\documentclass[prc,floatfix,nofootinbib,showpacs,showkeys,reprint]{revtex4-1}
\usepackage[latin1]{inputenc}
\usepackage{amsmath}
\usepackage{amsfonts}
\usepackage{amssymb}
\usepackage{graphicx}
\usepackage{url}

\begin{document}

\title{Cross Sections for A(e,e')X Reactions}

\author{U. Mosel}
	\email[Contact e-mail: ]{mosel@physik.uni-giessen.de}
	\affiliation{Institut f\"ur Theoretische Physik, Universit\"at Giessen, Giessen, Germany}
	\author{K. Gallmeister}
	\affiliation{Institut f\"ur Theoretische Physik,\\ Johann Wolfgang Goethe-Universit\"at Frankfurt, Frankfurt a.\ M., Germany}

\begin{abstract}
Recent JLAB measurements of inclusive (e,A) cross sections as a function of outgoing lepton momentum may provide information on the nucleon spectral function in targets that will be relevant for the planned DUNE neutrino long-baseline experiment. They also provide an important testing ground for neutrino generators. We have therefore used the transport theoretical framework and code GiBUU to calculate the cross sections for the targets C, Ar and Ti. We compare the calculations with the experimental data. The overall agreement is good. Relatively small discrepancies appear mainly on the low-electronenergy side of the QE-peak where QE and 2p2h excitation processes overlap. 
\end{abstract}

\maketitle

\section{Introduction}
Among the top priorities in particle physics are the precise determination of neutrino mixing angles, of the mass-ordering and of the CP symmetry violation in weak interactions. The experiment DUNE (Deep Underground Neutrino Experiment), with a beam originating at Fermilab and a far-detector located deep underground at the Sanford Lab, about 1300 km away, will be at the forefront of such studies for the coming two decades \cite{Acciarri:2015uup}.

Crucial for the extraction of the aforementioned properties is the knowledge of the neutrino energy since it enters explicitly into the oscillation formulas. As a consequence, the accuracy with which the incoming neutrino energy is known determines the accuracy of the extracted neutrino properties. Unlike in any other experiment in high-energy or nuclear physics the neutrino beam energy is not sharp, but is known only within a fairly broad distribution. At DUNE, e.g., the energy distributions peaks at about 3.5 GeV, with tails out to 30 GeV.

The neutrino energy is thus not fixed by some accelerator but must be obtained from an -- often incomplete -- observation of the final state of the neutrino-nucleus reaction. The reconstruction of the neutrino energy amounts to 'calculating backwards' from an observed final state. This is the purpose of generators such as GENIE, NEUT or NuWro \cite{Gallagher:2018pdg}. These generators must be able to handle not only the very first, initial interaction of the incoming neutrino with bound and Fermi-moving nucleons, but also the often sizeable final state interactions that the hadrons, produced in the first reaction, experience. The quality and predictive power of the 'backwards calculation' then depends critically on the quality of the underlying nuclear theory encoded in the generators \cite{Mosel:2016cwa}.

A crucial test for the quality of generators is provided by the description of electron interactions with nuclei since these are sensitive to both the vector part of the initial electroweak interaction and the same final state interactions. Although the neutrino experiments require the knowledge of the final state of the reaction, i.e. the four-vectors of all particles present, for the energy reconstruction a necessary test is already provided by inclusive cross sections. The new experimental data from Jefferson Laboratory, in which the inclusive cross sections were measured on C and Ti \cite{Dai:2018xhi} and Ar \cite{Dai:2018gch} at an energy of 2.222 GeV and a scattering angle of 15.541 degrees, provide such a testing ground. So far, no theoretical descriptions of all of these data were attempted; for C a calculation using the spectral function (SF) method is shown in \cite{Dai:2018xhi} and for Ar the results of a calculation using the Relativistic Green's Function (RGF) method only for the QE peak are given in \cite{Dai:2018gch}. Generator comparisons with these data are not available at all.

It is, therefore, the purpose of the present short paper to describe the results obtained with GiBUU, a theory and code framework for the description of nuclear interactions with nuclei \cite{Buss:2011mx}.

\section{Method}
We use GiBUU for the calculation of these inclusive cross sections as a function of outgoing lepton energy, i.e.\ essentially of the energy transfer. Unlike in neutrino experiments, where this energy transfer is not directly measurable, in electron experiments the energy-transfer is an observable. GiBUU cannot only provide the inclusive cross sections discussed in this paper, but it also describes the full time-development of a nuclear reaction, giving at the end the four-vectors of all particles present in the final state. GiBUU can, therefore, also be used as a generator.

We stress that all the subprocesses described in the following subsections consistently involve the same groundstate of the target nucleus (with the exception of the 2p2h process where we rely on a phenomenological description). This consistency is not present in other generators; there pieces from different theories are used for the description of the various interactions. This combination of various ingredients in the description of one and the same nucleus introduces artificial degrees of freedom.

For details we refer to the GiBUU review which contains all the theoretical basis and information on the technical implementation \cite{Buss:2011mx}\footnote{The code can be downloaded from {\it gibuu.hepforge.org}}. For the results to be discussed in the following section no special tunes were used; they were obtained 'out of the box'.

\subsection{Ground State Potential}
The GiBUU theory prepares the ground state by first calculating for a given realistic density distribution a mean field potential from a density- and momentum-dependent energy-density functional, originally proposed for the description of heavy-ion reactions \cite{Welke:1988zz}.
The potential obtained from it is given by
\begin{equation}  \label{U(p)}
U[\rho,p] = A \frac{\rho}{\rho_0} + B \left(\frac{\rho}{\rho_0}\right)^\tau + 2 \frac{C}{\rho_0} \int d^3p' \, \frac{f(\vec{r},\vec{p}')}{1 + \left(\frac{\vec{p} - \vec{p}'}{\Lambda}\right)^2}
\end{equation}
which is explicitly momentum dependent. Here $\rho_0$ is the nuclear equilibrium density and $f(\vec{r},\vec{p}')$ is the nuclear phase-space density obtained from the local Fermi-gas model
\begin{equation}
f(\vec{r},\vec{p}') = \frac{4}{(2\pi)^3} \Theta[(|\vec{p}| - p_F(\vec{r})] \quad {\rm with} \quad p_F = \left(\frac{3}{2} \pi^2 \rho(\vec{r})\right)^{1/3} ~.
\end{equation}
This same phase-space distribution and the momentum-dependent potential also enter into the calculation of matrix elements and transition probabilities. By an iterative procedure we make sure that the Fermi-energy is constant over the nuclear volume; the binding is fixed to -8 MeV for all nuclei.

\begin{figure}
\includegraphics[width=0.45 \textwidth]{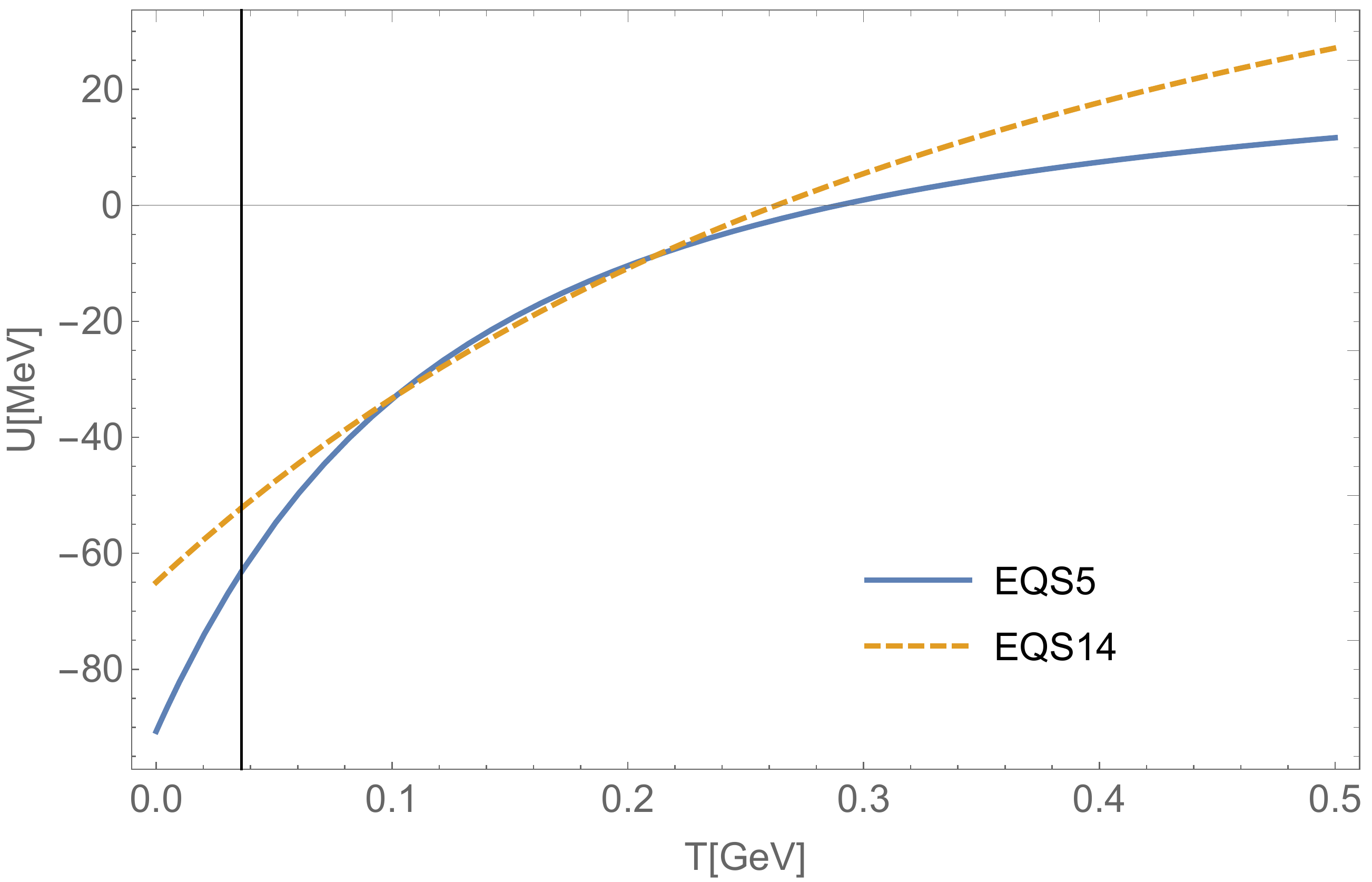}
\caption{The two potentials corresponding to the parametersets EQS5 (solid) and EQS14 (dashed) at density $\rho_0=0.16 fm^{-3}$ as a function of kinetic energy. EQS14 (dashed) is obtained from a fit to experimental pA data \cite{Cooper:1993nx} at $r=0$.}
\label{fig:pot}
\end{figure}

The parameters appearing in this potential are given for two different parameter sets in Table \ref{Parms}. Here the parameterset labeled EQS5 is that used in all our former calculations of neutrino interactions with nuclei; it was originally obtained from a fit to heavy-ion data \cite{Teis:1996kx}. The other parameterset EQS14 approximates the potential obtained by Cooper et al from a fit to p+A data \cite{Cooper:1993nx}.
\begin{table}

\begin{tabular}{|c|c|c|c|c|c|}
\hline
 EQS& A [MeV] & B [MeV] & $\tau$ &C [MeV]  & $\Lambda$ [fm$^{-1}$] \\
\hline
EQS5  & -29.25 & 57.25 & 1.76 & -63.52  & 2.13 \\
EQS14 & -124.8 & 204.8 & 1.2  & -77.29  & 4.0 \\
\hline
\end{tabular}
\caption{The parameters for the baryon potential of Eq.\ \ref{U(p)} for two different parameterizations. EQS14 describes the Schroedinger-equivalent potential extracted by Cooper et al from p+A scattering data \cite{Cooper:1993nx}}
\label{Parms}

\end{table}
The two potentials are shown in Fig.\ \ref{fig:pot} in their dependence on momentum $p$.
The potentials EQS5 and EQS14 agree with each other in the kinetic energy range $100 < T < 300$ MeV. For lower kinetic energies, corresponding to lower energy transfers, EQS5 gives a better description of the data (see the results shown in \cite{Gallmeister:2016dnq}). 

This very same potential is present both for the initial interaction as well as for the final state propagation of outgoing baryons. There is thus no need to invent artificial binding energy parameters as usually used in all other generators which do not contain any nuclear binding \cite{Bodek:2018lmc}. We note that none of the other generators have a potential for the outgoing particles implemented. In GENIE and NuWro a potential for the final state phase space of the first, initial interaction can be added 'by hand'. However, since their final state propagation does not allow for any potential this creates unphysical potential steps with corresponding spikes in the forces. In the SF method the final state potential is added by hand to the initial process \cite{Ankowski:2014yfa} by correcting only the final state phase-space. There is no consistency with the initial state potential that is effectively hidden in the SF.

\subsection{Inelastic Excitations}
Inelastic reactions in GiBUU are described by the explicit treatment of nucleon resonances (for an invariant mass $W < 2$ GeV) and their decay, based on the MAID07 analysis, and by DIS processes, as described by PYTHIA, for higher $W$ \cite{Leitner:2008ue}.

\subsection{2p2h Interactions}
In lepton interactions with nuclei also so-called 2p2h processes can take place \cite{Alberico:1983zg,Delorme:1985ps}. Since there is no theoretical framework available to calculate these processes starting from the GiBUU ground state and for the kinematical regime relevant for DUNE we take this 2p2h contribution from a global fit to electron scattering data on carbon over a wide kinematical range; for details see \cite{Gallmeister:2016dnq}. Since the MEC contributions are rather short-ranged we take them directly proportional to the nuclear mass number $A$.

\subsection{Final State Interactions}
For inclusive cross sections final state interactions play a role only through the final states of the very first lepton-nucleus interaction. We obtain these from the very same potential as the one used for the construction of the ground state, as explained above in Eq.\ (\ref{U(p)}).

\section{Results}
\begin{figure}
\includegraphics[width=0.45 \textwidth]{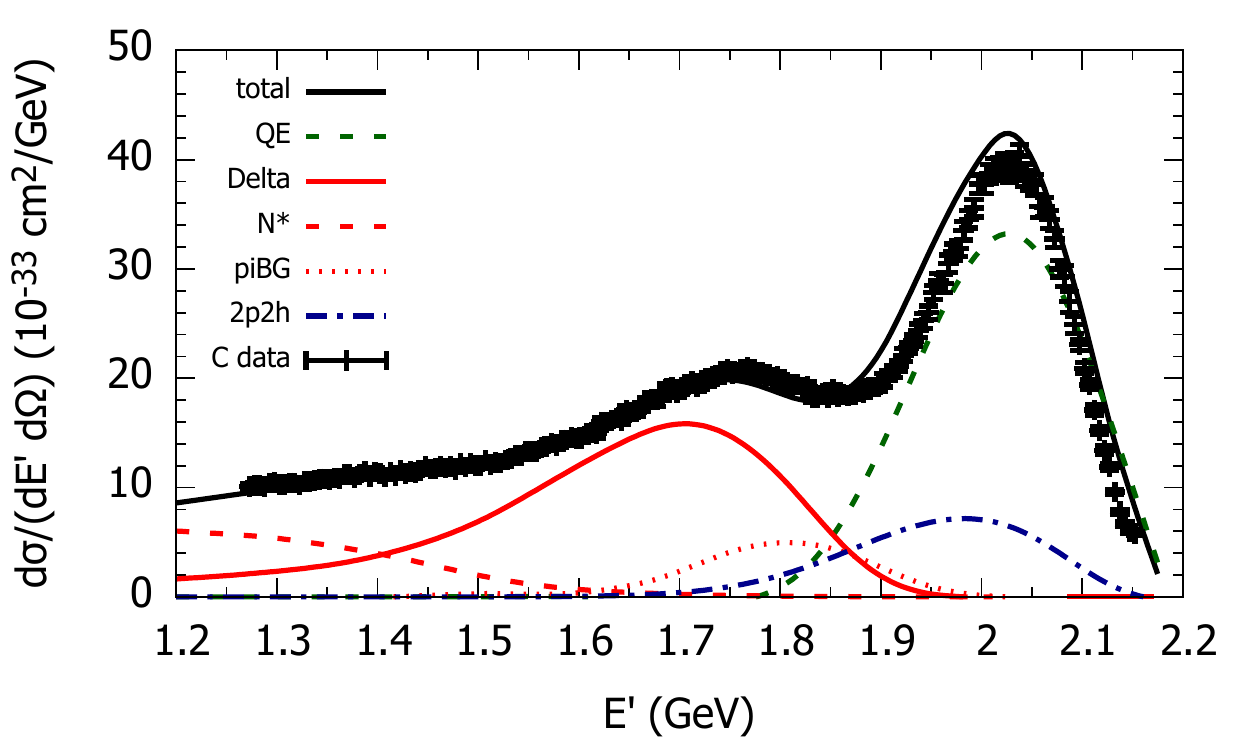}
\caption{Inclusive cross section for (e,$^{12}$C) at 2.222 GeV and 15.541 deg as a function of outgoing electron energy $E'$. The various curves give the contributions of different interaction processes to the total.  The data are taken from Ref.\ \cite{Dai:2018xhi}.}
\label{fig:C}
\end{figure}

In Figs. \ref{fig:C}, \ref{fig:Ar} and \ref{fig:Ti} we show the inclusive cross section for the 3 targets $^{12}$C, $^{40}$Ar and $^{48}$Ti as a function of outgoing electron energy. A first look shows that the agreement is quite good for all 3 targets over the full kinematical range, as far as the general structure and the overall magnitude of the cross section are concerned.

\subsection{Comparison with experiment}

In all three cases the agreement in the resonance region is perfect. In the QE region, however, a discrepancy on the left (lower electron energy) side of the QE peak shows up. There the theory gives too much strength for energy transfers of about 200 - 300 MeV. The results of a spectral function method shown in Ref.\ \cite{Dai:2018xhi} for $^{12}$C show a better description of the QE peak, but worse agreement both in the dip region and the high energy transfer region. The results of a RGF calculation, shown in Ref.\ \cite{Dai:2018gch} for Ar, describe the QE peak very well, but do not contain any inelastic components.

\begin{figure}
\includegraphics[width=0.45 \textwidth]{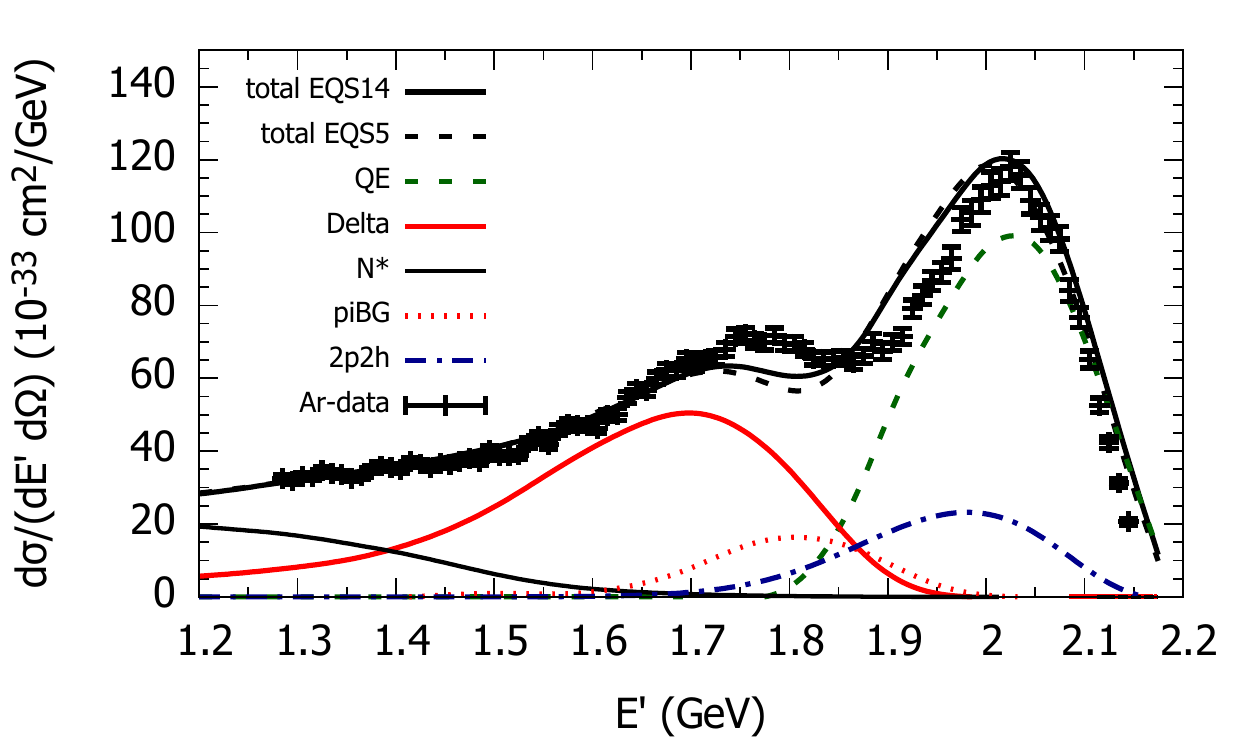}
\caption{Inclusive cross section for (e,$^{40}$Ar) at 2.222 GeV and 15.541 deg as a function of outgoing electron energy $E'$. The various curves give the contributions of different interaction processes to the total. The black solid line gives the total cross section calculated with EQS14, the topmost dashed black line, which nearly overlaps with the solid line, gives the total cross section calculated with EQS5. The data are taken from Ref.\ \cite{Dai:2018gch}.}
\label{fig:Ar}
\end{figure}
\begin{figure}
\includegraphics[width=0.45 \textwidth]{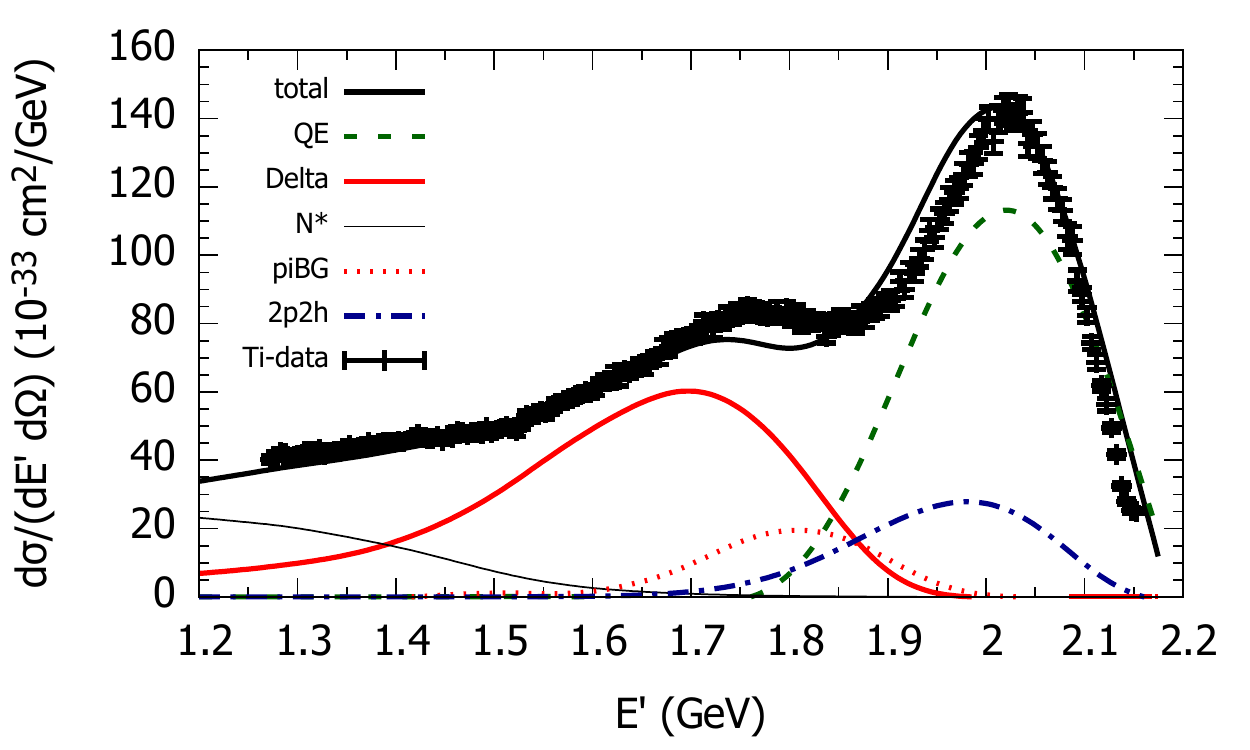}
\caption{Inclusive cross section for (e,$^{48}$Ti) at 2.222 GeV and 15.541 deg as a function of outgoing electron energy $E'$. The various curves give the contributions of different interaction processes to the total. The data are taken from Ref.\ \cite{Dai:2018xhi}.}
\label{fig:Ti}
\end{figure}

We now discuss the observed discrepancy at the high-energy-transfer side of the QE peak. Its origin is nevertheless not so clear since in this region the QE contribution overlaps mainly with the 2p2h contribution and -- to a lesser extent -- also with that of the $\Delta$ excitation and the pion background component. We note, first, that the agreement with data could be somewhat improved by shifting the energy-transfer by only about 20 MeV. This shift could be achieved by changing the momentum dependence of the single particle potential. While in a QE scattering process the initial nucleon has a low kinetic energy below the Fermi-energy, after the energy transfer of $\approx 350$ MeV it is close to 400 MeV and thus sees a much less attractive potential (see Fig.\ \ref{fig:pot}). Comparing the uppermost solid (EQS14) and dashed (EQS5) curves in Fig.\ \ref{fig:Ar} shows that the sensitivity of the calculated results to these equation of state is very weak, because both potentials agree in the relevant kinetic energy range.

Furthermore, it is not clear if the observed discrepancy is due to the QE process alone or whether the 2p2h contribution, which is quite significant in the lower-energy region of the QE peak, also contributes to the observed disagreement.  We have, therefore, abstained from tuning the potential to these kinematics since within the potential form of Eq.\ (\ref{U(p)}) and the constraints by the Cooper potential it seems to be difficult to maintain simultaneously the good agreement at the lower energies obtained in \cite{Gallmeister:2016dnq}. 

Anyway, since the overall energy transfer here is about 350 MeV, the necessary correction amounts to only about 5 \% of the energy transfer. This relatively small shift will play an even smaller role in neutrino-induced reactions where the energy transfer is smeared over because of the width of the incoming neutrino energy distribution. Furthermore, at the planned DUNE experiment the incoming beam energy is higher than here so that DIS will play a larger role. It is, therefore, encouraging to see that the present calculations give perfect agreement in particular in the region of large energy transfers.

\subsubsection{Z-Scaling}
\begin{figure}
	\includegraphics[width=0.45 \textwidth]{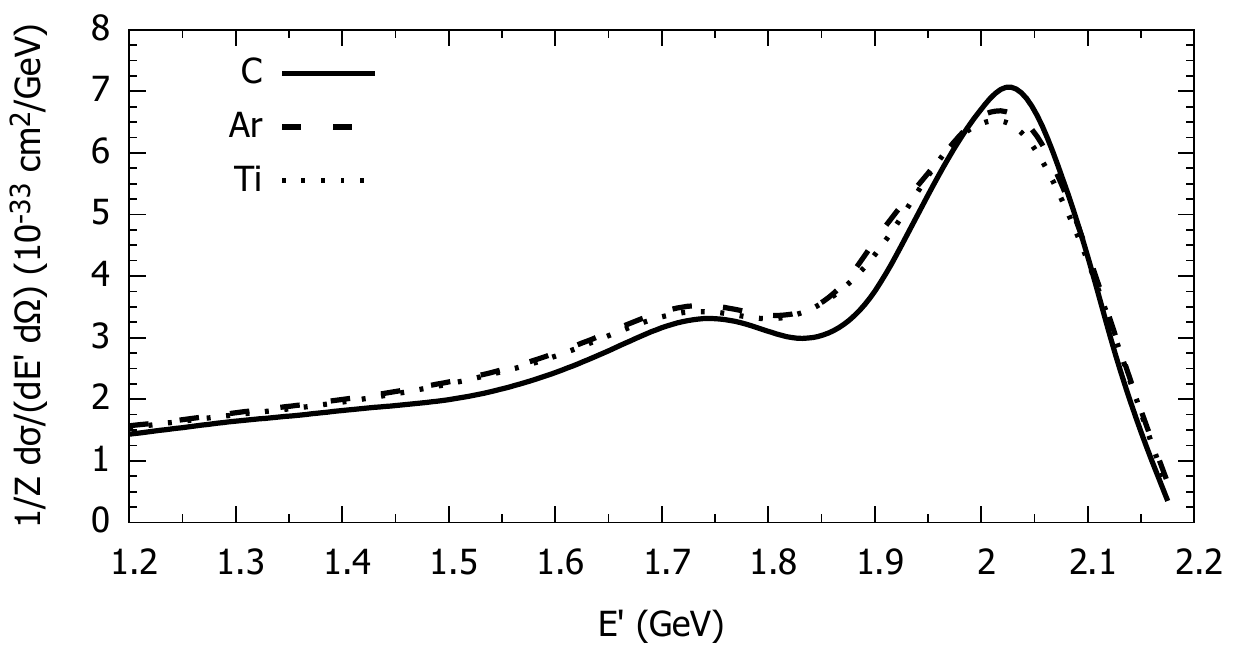}
	\caption{Inclusive cross section per proton for (e,C), (e,Ar) and (e,Ti) at 2.222 GeV and 15.541 deg as a function of outgoing electron energy $E'$.}
	\label{fig:Z-comp}
\end{figure}

Fig.\ \ref{fig:Z-comp} shows a comparison of the total cross sections for all 3 targets normalized to the target proton number. The scaled cross sections are very close to each other for all energies. 
This scaling with $Z$ reflects the fact that all reaction processes (QE, resonance excitations and 2p2h) are all one-particle or short-ranged processes. The curves for Ar and Ti are on top of each other while the curve for C is somewhat narrower and higher in the QE peak; this same behavior is also seen in the data (Fig. 2 in \cite{Dai:2018gch}).

\section{Summary and Conclusions}
In this paper we have compared the results of GiBUU calculations with very recent inclusive electron scattering data on nuclei that were taken with an aim to provide a basis and crucial input for neutrino generators. The comparison was done without any tuning and good agreement was obtained. A persistent discrepancy at the high energy-transfer side of the QE peak could possibly be traced back to the momentum-dependence of the single particle potential and/or to the behavior of the 2p2h contribution that extends into the region of the QE peak. In all three targets there is a significant 2p2h strength already at energies below the QE peak; this will complicate the extraction of the single-particle spectral function, one of the original aims of the JLAB experiment.

Since the description of electron scattering data on nuclei is usually seen as a crucial test for generators the comparison presented here is a first step into that direction (see also the lower-energy comparisons in Ref.\ \cite{Gallmeister:2016dnq}).  It would be instructive to see also results of similar comparisons of the data with the results of other generators as a first, necessary test of their reliability.

\section{Note added in proof}
After this paper was uploaded to the arXiv we became aware of results obtained with the GENIE generator by A. Ankowski and A. Friedland \cite{Ankowski:2019}. Also, results of a calculation using the SUSA scaling model have become available in the meantime \cite{Barbaro:2019vsr}.
\begin{acknowledgments}
We thank A. Ankowski and V. Pandey for providing us with the data points of the JLAB experiment.
\end{acknowledgments}

\bibliographystyle{apsrev4-1}
\bibliography{nuclear}

\end{document}